\documentclass[12pt,preprint]{aastex}

\slugcomment{Accepted by AJ on 2015 July 27}

\shorttitle{Intermediate-Mass Black Holes in Star Clusters}
\shortauthors{Wrobel, Nyland, \& Miller-Jones}

\begin{document}

\title{A Stacked Search for Intermediate-Mass Black Holes in 337
  Extragalactic Star Clusters}

\author{J. M. Wrobel\altaffilmark{1,}\altaffilmark{2},
  K. E. Nyland\altaffilmark{3}, and
  J. C. A. Miller-Jones\altaffilmark{4}}

\altaffiltext{1}{National Radio Astronomy Observatory, P.O. Box O,
  Socorro, NM 87801, USA; jwrobel@nrao.edu}

\altaffiltext{2}{The National Radio Astronomy Observatory (NRAO) is a
  facility of the National Science Foundation, operated under
  cooperative agreement by Associated Universities, Inc.}

\altaffiltext{3}{Netherlands Institute for Radio Astronomy, Postbus 2,
  7990 AA, Dwingeloo, The Netherlands; nyland@astron.nl}

\altaffiltext{4}{International Centre for Radio Astronomy Research,
  Curtin University, GPO Box U1987, Perth, WA 6845, Australia;
  james.miller-jones@curtin.edu.au}

\begin{abstract}
Forbes et al.\ recently used the {\em Hubble Space Telescope\/} to
localize hundreds of candidate star clusters in NGC\,1023, an
early-type galaxy at a distance of 11.1 Mpc.  Old stars dominate the
light of 92\% of the clusters and intermediate-age stars dominate the
light of the remaining 8\%.  Theory predicts that clusters with such
ages can host intermediate-mass black holes (IMBHs) with masses
$M_{\rm BH} \lesssim 10^5~M_\odot$.  To investigate this prediction,
we used 264 s of 5.5 GHz data from the Karl G.\ Jansky Very Large
Array (VLA) to search for the radiative signatures of IMBH accretion
from 337 candidate clusters in an image spanning 492\arcsec\, (26 kpc)
with a resolution of 0\farcs40 (22 pc).  None of the individual
clusters are detected, nor are weighted-mean image stacks of the 311
old clusters, the 26 intermediate-age clusters, and the 20 clusters
with stellar masses $M_\star \gtrsim 7.5 \times 10^5~M_\odot$.  The
clusters thus lack radio analogs of HLX-1, a strong IMBH candidate in
a cluster in the early-type galaxy ESO\,243-49.  This suggests that
HLX-1 is accreting gas related to its cluster's light-dominating young
stars.  Alternatively, the HLX-1 phenomenon could be so rare that no
radio analog is expected in NGC\,1023.  Also, using a formalism
heretofore applied to star clusters in the Milky Way, the
radio-luminosity upper limit for the massive-cluster stack corresponds
to a 3$\sigma$ IMBH mass of $\overline{M_{\rm BH}({\rm massive})} <
2.3 \times 10^5~M_\odot$, suggesting black-hole mass fractions of
$\overline{M_{\rm BH}({\rm massive})} / M_\star < 0.05-0.29$.
\end{abstract}


\section{Motivation}\label{motivation}

Constraints on populations of low-mass black holes (BHs) are important
for understanding formation channels for seed BHs in the early
Universe and the demographics of relic seed BHs in the local Universe
\citep[reviewed by][]{gre12,vol12}.  Such seeds are referred to as
intermediate-mass black holes (IMBHs) because they occupy the gap in
mass between the well-studied stellar-mass BHs with $M_{\rm BH} \sim
10~M_\odot$ and the well-established supermassive BHs with $M_{\rm BH}
\gtrsim 10^6~M_\odot$.  In addition, if the scaling relations between
central BHs and their host stellar systems differ between the
intermediate-mass and supermassive ends, then this could provide clues
about the co-evolution, or not, of these entities
\citep[e.g.,][]{xia11,jia11,kor13}.

Of particular interest here are the theoretical predictions that dense
star clusters could host IMBHs formed either slowly through the
successive mergers of stellar-mass BHs \citep{mil02} or rapidly
through the runaway collisions of massive stars in the cluster centers
\citep{gur04,por04}.  In addition, if IMBHs exist in star clusters,
the inspiral of stellar-mass BHs into IMBHs is expected to be a key
source of gravitational waves for planned space missions
\citep[e.g.,][]{kon13}.  Overall, finding IMBHs in star clusters could
validate a theoretical formation channel, bolster gravitational-wave
predictions, and test scaling relations between stellar systems and
the central BHs they host.

At distances of about 10 Mpc, future near-infrared telescopes will be
able to measure BH masses as low as $M_{\rm BH} \sim 10^5~M_\odot$ by
resolving their spheres of influence in their host stellar systems
\citep{do14}.  While this is expected to lead to a robust inventory of
IMBHs in the local Universe, these next-generation facilities are
still many years off.  Moreover, such sphere-of-influence studies have
already been applied to star clusters in the Local Group, and all
these studies are controversial \citep[][and references
  therein]{str12a}.

Due to these on-going controversies and not wanting to wait for future
facilities, we are pursuing an independent route, namely searching for
radio signatures of accretion onto putative IMBHs in extragalactic
star clusters \citep{mac04}.  Analogies with stellar-mass BHs suggest
that an IMBH will spend more time in the hard X-ray state associated
with a low accretion rate, than in the soft X-ray state associated
with a high accretion rate \citep[reviewed by][]{fen12}.  If only a
few radio observations are available, they are more likely to sample
the steady radio emission associated with the hard X-ray state, as
compared to the flaring radio emssion associated with a transition
from the hard X-ray state to the soft X-ray state.  We thus take three
approaches, as follows:

\begin{enumerate}
\item Radio emission resembling that from HLX-1 in its hard X-ray
  state \citep{cse15} could be directly detected from an individual
  cluster.
\item Using the empirical fundamental-plane regression for the hard
  X-ray state, constraints on X-ray and radio luminosities can yield
  an estimate of the mass of an IMBH \citep{mer03,fal04,plo12}.
\item A conservative, semi-empirical model can be used to predict the
  mass of an IMBH that, if undergoing Bondi accretion in the hard
  X-ray state, is consistent with the radio luminosity
  \citep{mac08,mac10,str12a}.
\end{enumerate}

In this paper, we use the Karl G.\ Jansky Very Large Array
\citep[VLA;][]{per11} to search for radio emission from hundreds of
candidate star clusters localized by \citet{for14} in a mosaic of
\object[NGC1023]{NGC\,1023} with the Advanced Camera for Surveys (ACS)
on the {\em Hubble Space Telescope\/} (HST).  We focus on NGC\,1023
because it is one of only a few early-type galaxies with such an
extensive ACS mosaic \citep{for14} and, at a distance of 11.1 Mpc, is
amongst the nearest galaxies in the extensive VLA survey of K.\ Nyland
et al.\ (2015, in preparation).  At this distance 1\arcsec\, = 53.8
pc.

After folding in prior information
\citep{lar00,lar01,lar02,bro02,mor06,chi13}, \citet{for14} divide
their candidate star clusters by age: stars older than 5 Gyr dominate
the light of 92\% of the clusters, whereas stars with ages of a few
hundred Myr dominate the light of the remaining 8\%.  Those latter
intermediate-age clusters may have formed during a recent gas-rich
interaction between NGC\,1023 and its companion NGC\,1023A, with a
projected offset of $2\farcm5$ (8 kpc).  Still, the vast majority of
the clusters are dominated by an old, globular-cluster-like stellar
population.  Using that population's z-band (F850LP) mass-to-light
ratio from \citet{jor07} and magnitudes from \citet{for14}, the
stellar masses of the clusters range from $M_\star \sim 5 \times
10^4~M_\odot$ to $M_\star \sim 4 \times 10^6~M_\odot$.  It is also
important to underscore that only a handful of the candidate clusters
have had their association with NGC\,1023 or its companion confirmed
spectroscopically.  \citet{liu11} used Chandra data to localize
several dozen X-ray sources but did not attempt to match those sources
with candidate star clusters.

The new VLA results are described in \S~\ref{imaging}.  No individual
cluster is detected, and \S~\ref{implications} explores the
implications of these nondetections regarding HLX-1 analogs, X-ray
detected clusters and semi-empirical model predictions.
\S~\ref{sumcon} contains a summary and conclusions.

\section{Imaging}\label{imaging}

The A configuration of the VLA was used to observe NGC\,1023 on 2012
October 3 UT as part of an extensive survey for low-luminosity active
galactic nuclei in early-type galaxies (proposal code 12B-281, PI
K.\ Nyland).  Full details will appear in K.\ Nyland et al.\ (2015, in
preparation) but key aspects are given here.  The observations were
made assuming a coordinate equinox of 2000 and were phase referenced
to the calibrator J0230+4032 at an assumed position of $\alpha(J2000)
= 02^{h} 30^{m} 45\fs7108$ and $\delta(J2000) = 40\arcdeg 32\arcmin
53\farcs068$ with one-dimensional errors at 1$\sigma$ of 2 mas.  The
switching angle between NGC\,1023 and J0230+4032 was $2\fdg9$.  One
observation of NGC\,1023 was immediately preceeded and followed by
observations of J0130+0432, leading to a net exposure time on
NGC\,1023 of 264 s.  Every 1 s the correlator generated 1024
contiguous 2 MHz channels, resulting in a total bandwidth of 2.048 GHz
per circular polarization centered at a frequency of 5.5 GHz.
Observations of 3C\,48 were also acquired to set the amplitude scale
to an estimated accuracy of about 3\%.  All but one of the 27 antennas
provided data of acceptable quality.  Given the observing strategies,
the one-dimensional astrometric error at 1$\sigma$ is estimated to be
$0\farcs1$.

Release 4.1.0 of the Common Astronomy Software Applications (CASA)
package \citep{mcm07} was used to calibrate and edit the visibility
data in version 1.2.0 of an automated pipeline.\footnote{
  https://science.nrao.edu/facilities/vla/data-processing/pipeline}
The CASA task {\tt clean} was used to form and deconvolve an image of
the Stokes $I\/$ emission from NGC\,1023 that spanned $8000 \times
0\farcs075 = 10\arcmin$, well matched to the HST mosaic \citep{for14}.
The CASA task {\tt clean} was run with (1) a robustness parameter of
0.5 to obtain the best compromise among sensitivity, spatial
resolution and sidelobe suppression; (2) an nterms parameter of 2 to
account for the large fractional bandwidth; and (3) the gridmode
parameter set to ``widefield'' and the wprojplanes parameter set to
128 to correct for the effects of non-coplanar baselines.

The CASA task {\tt widebandpbcor} was then run to correct the VLA
image for primary beam attenuation, characterized by a half-power
diameter of $8\farcm2$ where the attenuation is 50\% and thus the
correction is 200\%.  For comparison, at that location the response to
a point source is estimated to be reduced by about 2\% due to
bandwidth smearing across 2 MHz channels and by less than 1\% due to
time averaging over 1 s integrations,\footnote{
  https://science.nrao.edu/facilities/vla/docs/manuals/oss/performance/fov}
and no corrections were made for these negligible effects.  A square
image spanning the half-power diameter of the primary beam on each
axis was used for subsequent analysis.

This VLA image at 5.5 GHz captured 339 star cluster candidates
identified and cataloged by \citet{for14}.  For each of those
clusters, the 2015 December 31 release of NRAO's Astronomical Image
Processing System \citep[AIPS;][]{gre03} was used to form a cutout,
with task {\tt subim}, spanning $8\arcsec$ and centered on the optical
position \citep{for14}.  The ACS images were astrometrized by matching
the extracted sources with CFHT archival catalogs with one-dimensional
errors at 1$\sigma$ estimated to be 0\farcs15 (V.\ Pota, 2014,
priv.\ comm.).  The cutout for cluster ID 39 in the nomenclature of
\citet{for14} was dropped from further analysis because it was
contaminated by a 5.5-GHz counterpart to a background QSO
\citep{lar02}.  The radio and optical positions of this background QSO
differed by $0\farcs2$, consistent with the error estimates quoted
above for the radio and optical systems.  ID 58 was located too close
to the edge of the VLA image to form a cutout and was also dropped
from further analysis.  After this culling, 337 cutouts remain for an
analysis of individual clusters (Figure 1).  The 1$\sigma$ rms noise
level amongst the cutouts has a range of 16-68~$\mu$Jy~beam$^{-1}$,
reflecting the primary beam attenuation.  To minimize the risk of a
false-positive detection of one or more individual clusters when
examining an ensemble of 337, we adopt an individual one-tail
detection threshold of 3.5$\sigma$ \citep{wal03}.

\begin{figure}[t]
\plotone{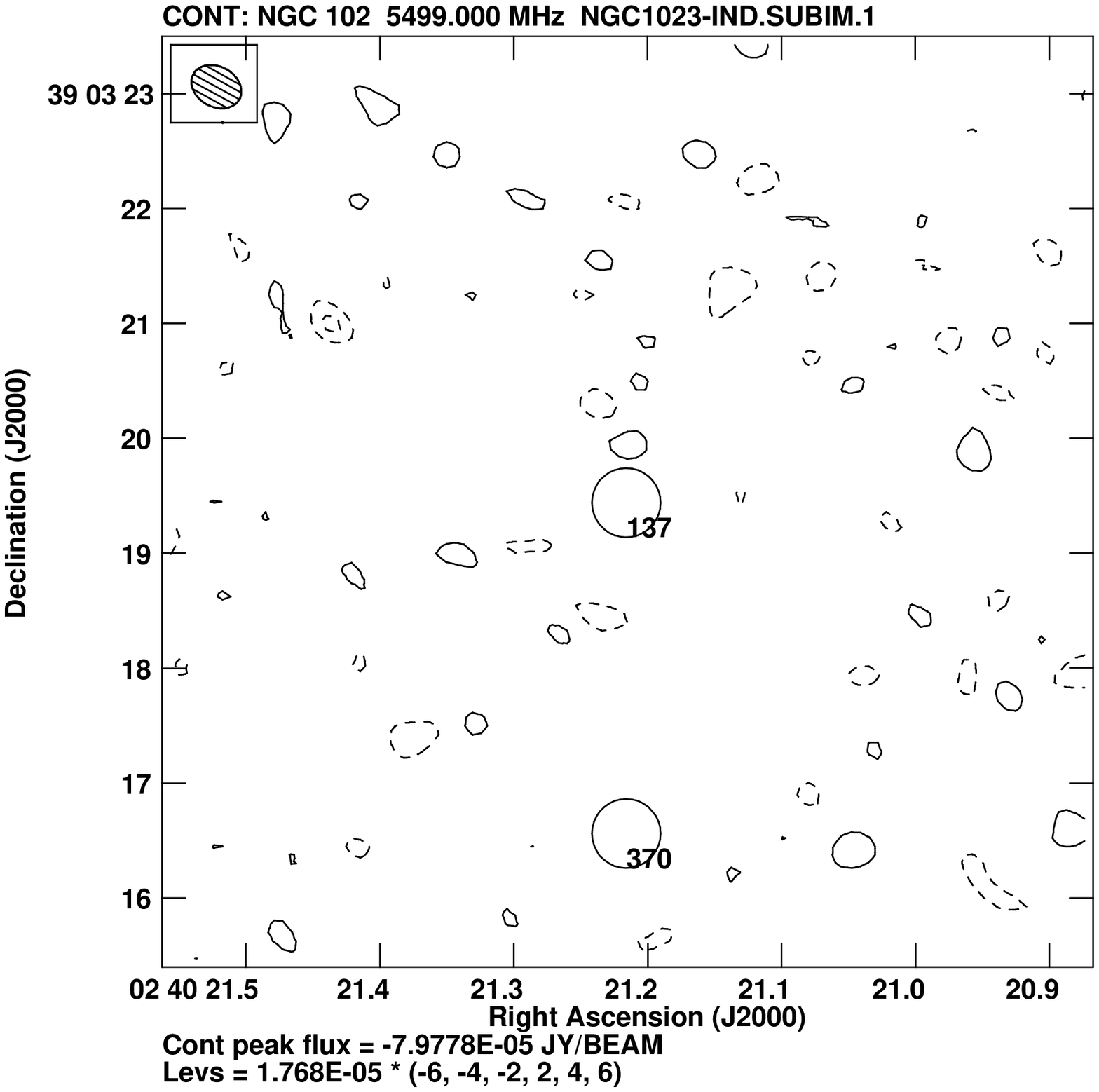}
\caption{VLA cutout of the Stokes $I\/$ emission at 5.5 GHz centered
  on the optical position of a candidate star cluster in NGC\,1023.
  The hatched ellipse in the north-east corner shows the beam
  dimensions at FWHM of 0\farcs46 (25 pc) $\times$ 0\farcs35 (19 pc)
  with an elongation position angle of 61\arcdeg.  Contours are at -6,
  -4, -2, 2, 4, and 6 times the 1$\sigma$ rms noise shown in the
  legend in units of Jy~beam$^{-1}$.  Linearly-spaced contours are
  optimal for conveying noise levels at a glance.  Negative contours
  are dashed and positive ones are solid.  The central circle of
  diameter 0\farcs6 (32 pc) shows the cluster's optical positional
  uncertainty at 95\% confidence and is labelled with its ID from
  Forbes et al. (2014).  The VLA photometry seeks evidence for the
  accretion signature of a point-like IMBH in the cluster's center.
  The cluster is not detected above the 3.5$\sigma$ level.  Figures
  1.1 $-$ 1.337 are available in the online version of the Journal.
  Some cutouts show additional clusters offset from the central
  cluster.  The 337 clusters have effective optical diameters of
  $0.8-40$ pc (Forbes et al. 2014) and their images are ordered by
  increasing cluster diameter.}\label{f1}
\end{figure}

\section{Implications}\label{implications}

None of the 337 candidate clusters in NGC\,1023 is detected above its
local 3.5$\sigma$ level at 5.5 GHz (Figure 1).  The corresponding
3.5$\sigma$ radio luminosities are $\nu L_\nu = L_{\rm R} < 4.5-19
\times 10^{34}$ erg s$^{-1}$ (Figure 2).  Such a luminosity definition
implicitly assumes a flat radio continuum spectrum up to 5.5 GHz.

\begin{figure}[t]
\includegraphics[angle=-90,scale=.65]{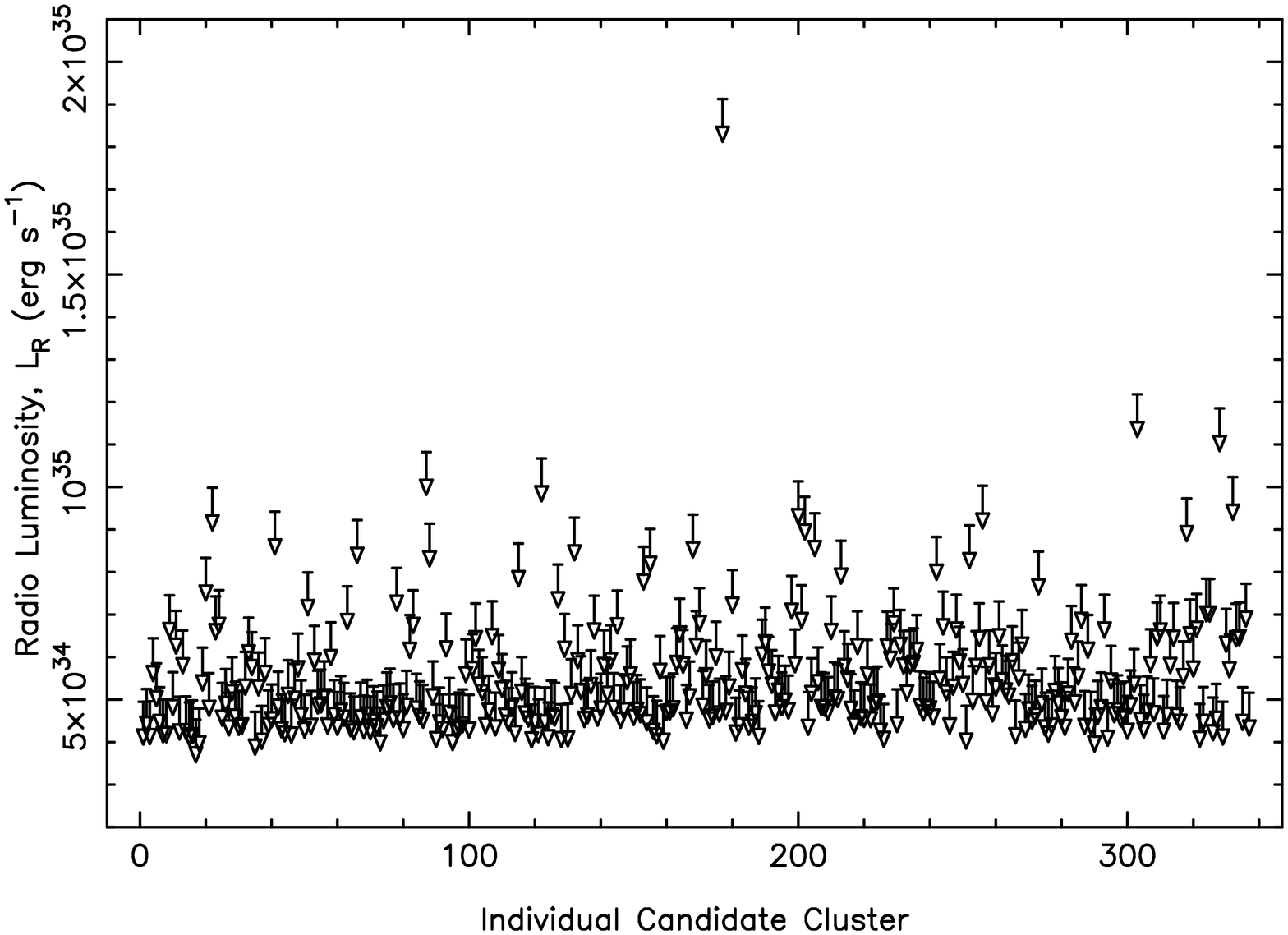}
\caption{Upper limits to the radio luminosities at 5.5 GHz, $L_{\rm
    R}$, of 337 individual candidate clusters in NGC\,1023.  The
  abscissa conveys the ordering of the clusters, 1 through 337, and
  corresponds to the ordering of Figures 1.1 - 1.337.}\label{f2}
\end{figure}

For the 311 old clusters, all accreting stellar-mass compact objects,
whether in the hard X-ray state \citep{str12b} or associated with a
transition from the hard X-ray state to the soft X-ray state
\citep{cor12}, would have radio luminosities too faint to be detected.
For the 26 intermediate-age clusters, no other data are available for
comparison.  However, \citet{kep14} showed that the massive star
clusters in II\,Zw\,40 at a distance of 10.5 Mpc have shed their
enshrouding nebular emission by an age of 10 Myr.  Specifically, the
10-Myr old cluster SSC-South in II\,Zw\,40 is not detected at 4.9 GHz
and has a radio-luminosity upper limit of $L_{\rm R} < 2.1 \times
10^{34}$ erg $s^{-1}$.  This strongly suggests that NGC\,1023's
ten-times-older clusters should lack contamination from free-free
nebular emission.  Thus, overall, we expect no contaminants for the
star clusters in Figure 1, allowing us to explore the implications of
the nondetections within the contexts of analogs of HLX-1
(\S~\ref{hlx}), X-ray detected clusters (\S~\ref{x-ray}), and
predictions of a semi-empirical model (\S~\ref{model}).

\subsection{HLX-1 Analogs}\label{hlx}

HLX-1 is a strong IMBH candidate with $M_{\rm BH} \sim
10^{4-5}~M_\odot$ that is hosted by a cluster with a stellar mass of
$M_\star \sim 10^{5-6}~M_\odot$
\citep{far09,sor10,wie10,far12,sor12,far14}.  The stellar mass of the
host cluster resembles those of the clusters in NGC\,1023, but the age
of the host cluster's light-dominating stars, about 20 Myr, is
significantly younger.

When transitioning from its hard to soft X-ray states, HLX-1 can
achieve a luminosity at 7 GHz of $L_{\rm R} \sim 3.4 \times 10^{36}$
erg s$^{-1}$ \citep{web12}.  Such a level is incompatible with the
5.5-GHz upper limits for individual star clusters in NGC\,1023 (Figure
2).

In its hard X-ray state, a combination of four measurements of HLX-1
yields a luminosity at 6.8 GHz of $L_{\rm R} \sim 1.6 \times 10^{36}$
erg s$^{-1}$, with no significant evidence for variability over
several days or months \citep{cse15}.  This steady luminosity is also
incompatible with the upper limits for individual star clusters in
NGC\,1023.  As \citet{cse15} argue, this steady emission from HLX-1 is
likely Doppler boosted by a factor of about five to ten, implying a
side-on luminosity of about $L_{\rm R} \sim 1.6-3.2 \times 10^{35}$
erg s$^{-1}$.  If such a side-on luminosity was present in any of the
337 individual clusters in NGC\,1023, Figure 2 shows that it could
escape detection in only a single cluster.

It is important to highlight the advantages of radio searches for
steady emission from HLX-1 analogs: its hyperluminous X-ray luminosity
of $L_{\rm X} \sim 10^{42}$ erg s$^{-1}$ is achieved only during a 2-3
week soft state that occurs approximately yearly \citep{god14} and can
thus be missed if only a few X-ray observations are made.  In
contrast, steady radio emission should be present during most of the
year between soft X-ray states, making it easier to detect even in a
single radio observation, the situation for NGC\,1023.  Also, at the
distance of NGC\,1023, the hyperluminous X-ray luminosity of HLX-1 is
below the sensitivity threshold of current and past X-ray all-sky
monitors, blocking those discovery routes.

Even more stringent constraints on HLX-1 analogs in NGC\,1023 come
from the weighted-mean image stacks \citep[e.g.,][]{lin15} in Figure 3
for the 311 old clusters and in Figure 4 for the 26 intermediate-age
clusters.  The stacks, which measure the mean contribution to the
total radio emission from each cluster population, were made using the
AIPS task {\tt comb}.  The stacks' 3$\sigma$ upper limits correspond
to radio luminosities of $\overline{L_{\rm R}({\rm old})} < 2.9 \times
10^{33}$ erg s$^{-1}$ and $\overline{L_{\rm R}({\rm
    intermediate~age})} < 1.2 \times 10^{34}$ erg s$^{-1}$.  These
stacks rule out emission from a population of HLX-1 analogs amongst
either cluster population.

\begin{figure}[t]
\plotone{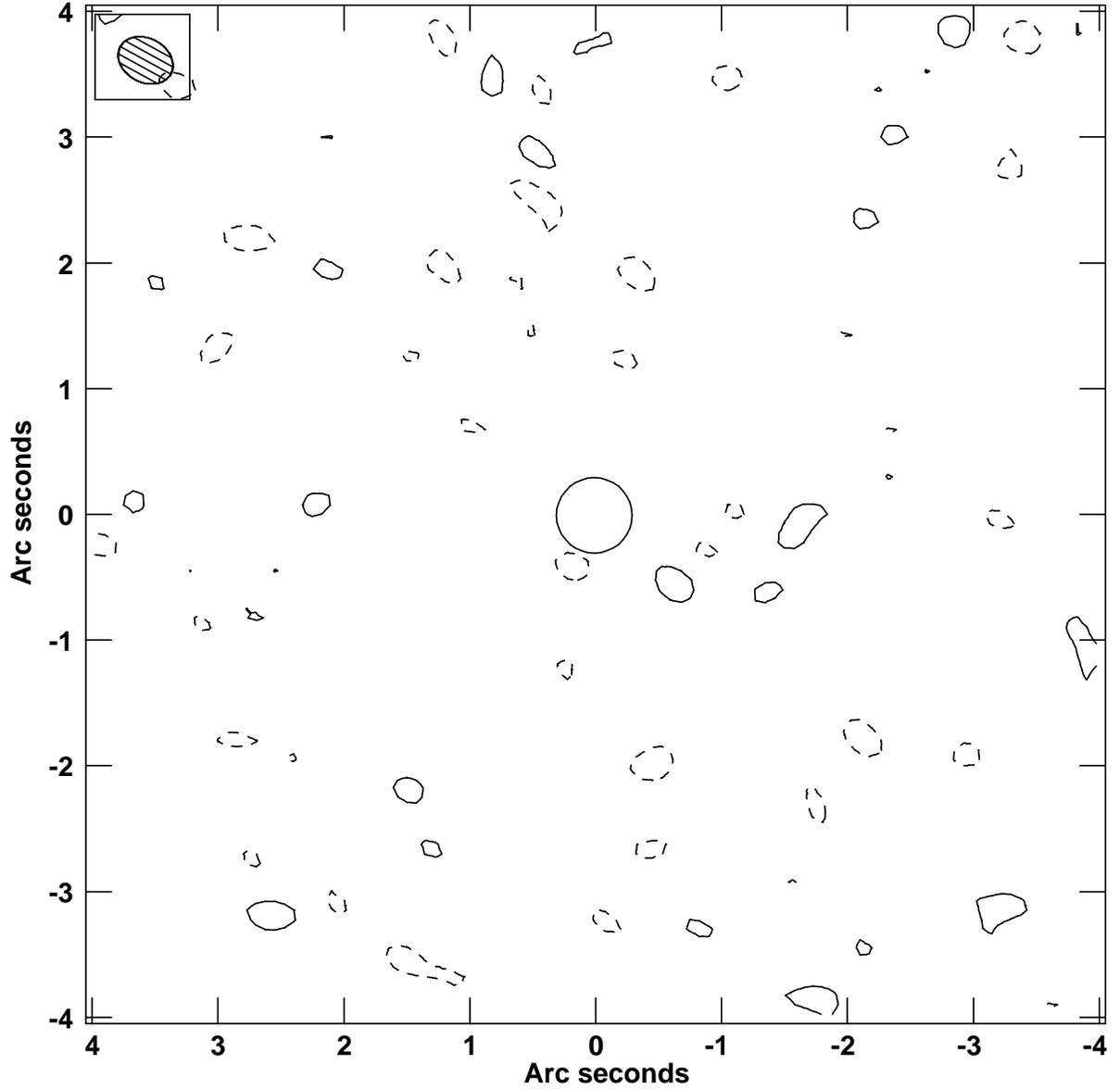}
\caption{Weighted-mean stack of the VLA images of the 311 old clusters
  in Figure 1.  The stacked image has an rms noise of
  1.2~$\mu$Jy~beam$^{-1}$ (1$\sigma$).  The hatched ellipse,
  contouring scheme and circle are the same as for Figure 1.  The
  stack measures the mean contribution to the total radio emission
  from the old clusters.  No emission is detected above 3$\sigma$ =
  3.6~$\mu$Jy~beam$^{-1}$.}\label{f3}
\end{figure}

\begin{figure}[t]
\plotone{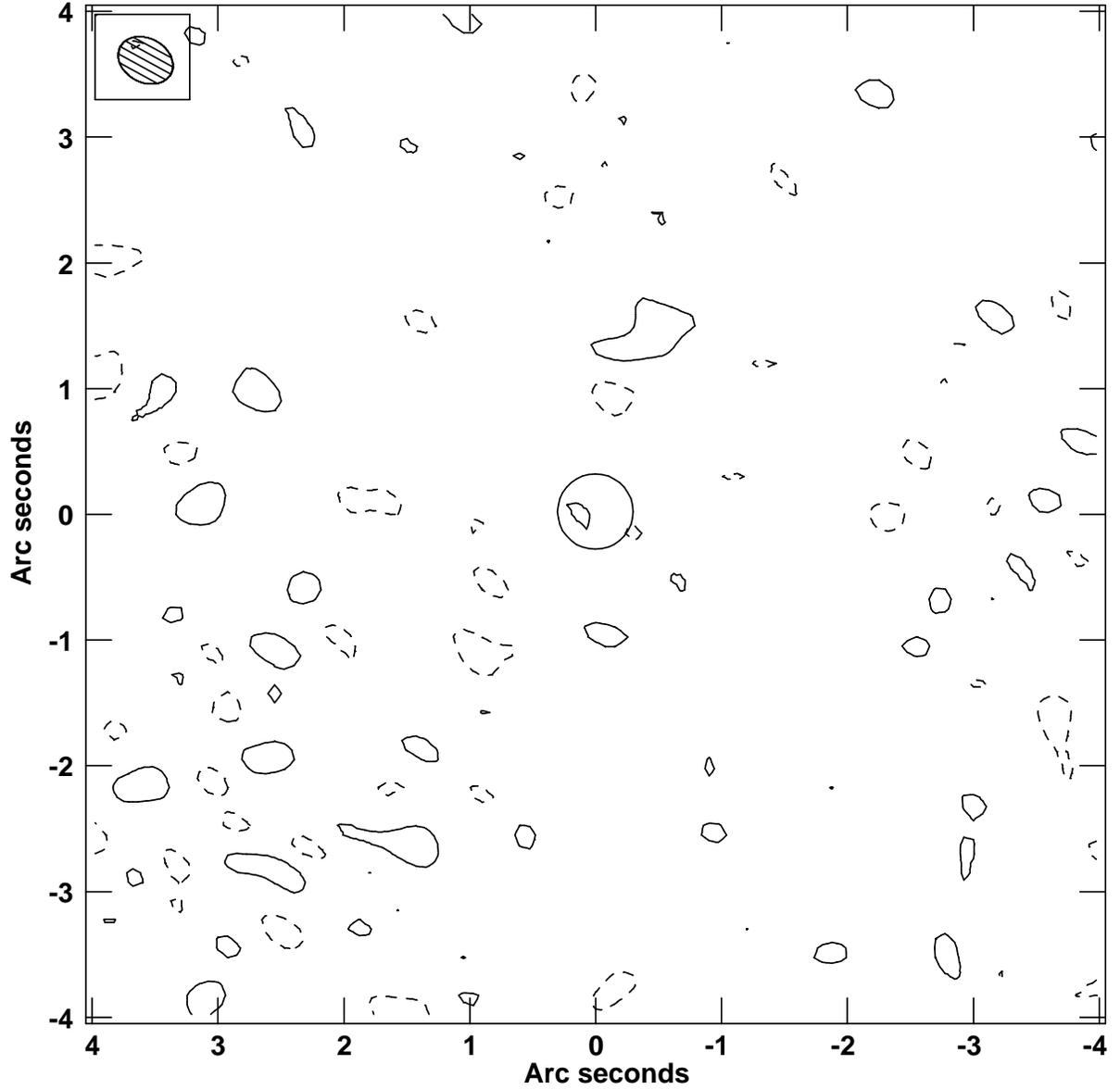}
\caption{Weighted-mean stack of the VLA images of the 26
  intermediate-age clusters in Figure 1.  The stacked image has an rms
  noise of 4.9~$\mu$Jy~beam$^{-1}$ (1$\sigma$).  The hatched ellipse,
  contouring scheme and circle are the same as for Figure 1.  The
  stack measures the mean contribution to the total radio emission
  from the intermediate-age clusters.  No emission is detected above
  3$\sigma$ = 14.7~$\mu$Jy~beam$^{-1}$.}\label{f4}
\end{figure}

The light-dominating stars in the clusters in NGC\,1023 are at least a
factor of ten older than the 20-Myr-old stars in the cluster hosting
HLX-1 \citep{far14}.  So one could speculate that if IMBHs exist in
NGC\,1023 clusters, those older clusters have lost their IMBHs due to
graviational wave recoil.  However, for an IMBH as massive as HLX-1,
$M_{\rm BH} \sim 10^{4-5}~M_\odot$, ejection from the cluster is very
improbable \citep{hol08}.  Instead, the absence of HLX-1 analogs in
NGC\,1023 suggests that HLX-1 is accreting gas somehow related to the
young stars in its host cluster.  An enhanced gas supply could arise
from material remaining after the burst of star formation 20 Myr ago,
or from material donated by a young stellar companion, as advocated by
\citet{mil14}.\footnote{20 Myr is approximately the main sequence
  lifetime of a 12 $M_\odot$ star \citep[e.g.,][]{for13}.}  Either
way, such an enhanced fuel supply would not be available to the IMBHs,
if any exist, in the star clusters in NGC\,1023.  Alternatively, it
might be that the HLX-1 phenomenon is so rare that a radio analog is
simply not expected in NGC\,1023.

\subsection{X-ray Detected Clusters}\label{x-ray}

Cluster IDs 153, 176 and 305 in the nomenclature of \citet{for14} have
X-ray detections, all of hard-state emission \citep{liu11}.  Using the
empirical fundamental-plane regression for the contracted sample of
\citet{plo12}, the clusters' X-ray luminosities from \citet{liu11} and
radio-luminosity upper limits from Figure 2 imply individual 3$\sigma$
IMBH masses of $M_{\rm BH} < 10^6~M_\odot$.  These clusters have old
stellar populations and estimated stellar masses of $M_\star \sim 1.3
\times 10^6~M_\odot$ or less.  But a star cluster is unlikely to host
an IMBH whose mass is comparable to, or exceeds, its stellar mass
\citep{mil02}, so these IMBH mass constraints are not physically
interesting.  Future imaging that reaches clusters with lower X-ray
luminosities would be especially beneficial.

\subsection{Semi-Empirical Model}\label{model}

Following \citet{mac08,mac10} and \citet{str12a}, we use a
semi-empirical model to predict the mass of a putative IMBH that, if
undergoing hard-X-ray-state accretion in a gas-poor cluster, is
consistent with the upper limit on the radio luminosity.
Specifically, we conservatively assume that the IMBH undergoes
accretion at a fraction $f_b = 0.03$ of the Bondi rate from a tenuous,
globular-cluster-like medium with a gas density $\rho = 0.2$
cm$^{-3}$.  The rationale for these assumptions and their
uncertainties are discussed in section 3.4 of \citet{str12a}.  These
parameters yield a prediction for the hard-state X-ray luminosity
$L_{\rm X}$.  The empirical fundamental-plane regression for the
contracted sample of \citet{plo12} is then used to predict the
associated radio luminosity $L_{\rm R}$.  (Using an earlier
regression, \citet{mac10} demonstrate that $M_{\rm BH} \propto L_{\rm
  R}^{0.38} (f_b \rho)^{-0.46}$.)  A detection or upper limit to a
radio luminosity thus maps to a detection or upper limit to an IMBH
mass.

The results of applying this semi-emipirical model to NGC\,1023 are
shown in Figure 5.  To avoid clutter only the best 3.5$\sigma$
constraint from Figure 2 for individual clusters in NGC\,1023 is
shown.  Figure 5 also conveys the 3$\sigma$ constraints from the
stacks of the old (Figure 3) and the intermediate-age (Figure 4)
clusters in NGC\,1023.  For comparison purposes, Figure 5 also shows
the best 3$\sigma$ constraints from uniformly applying the
semi-empirical model to individual clusters in the Milky Way
[\citet[][Pal\,2, NGC\,1851, NGC\,6440, NGC\,6624, NGC\,7099]{kna96};
  \citet[][NGC\,6397]{der06}; \citet[][M80]{bas08};
  \citet[][NGC\,2808]{mac08}; \citet[][NGC\,6388]{cse10};
  \citet[][M54]{wro11}; \citet[][47\,Tuc, $\omega$\,Cen]{lu11};
  \citet[][M15, M19, M22]{str12a}; \citet[][M62]{cho13}; using
  distances from \citet{har96}], as well as to the most luminous
cluster in M31 \citep[][G1]{mil12} using the distance from
\citet{vil10} and the most luminous cluster in M81
\citep[][GC1]{swa03,may13} using the distance from \citet{fre94}.

\begin{figure}[t]
\plotone{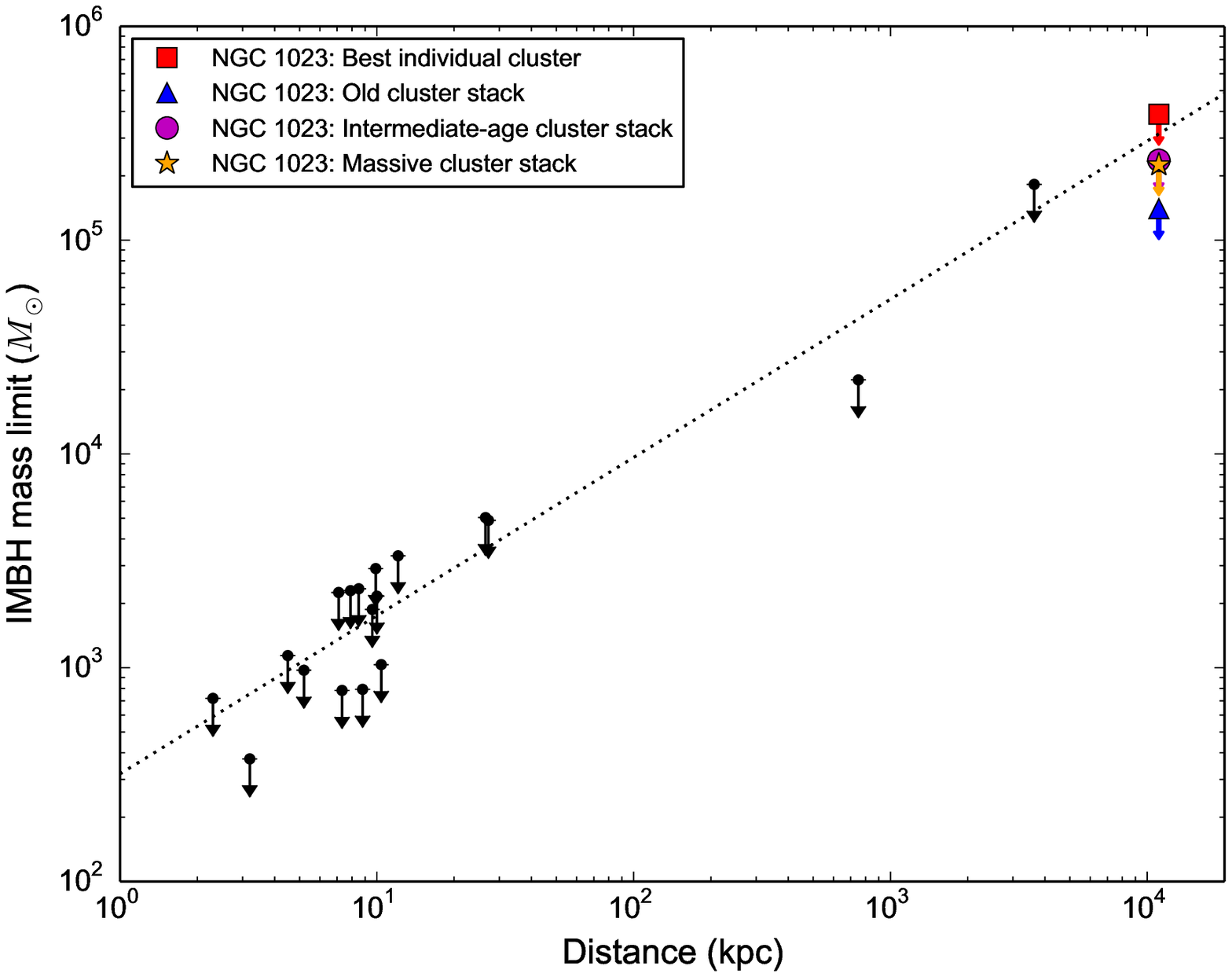}
\caption{IMBH masses predicted from a semi-empirical model for star
  clusters at various distances.  If a putative IMBH is undergoing
  hard-X-ray-state accretion in a gas-poor cluster, then a detection
  of, or upper limit to, its radio luminosity yields a value for, or
  upper limit to, its IMBH mass.  To visually convey the model's mass
  sensitivity with distance, the dotted line shows the trend for a
  fiducial frequency of 6 GHz and a fiducial flux density of
  30~$\mu$Jy.  Upper limits achieved for NGC\,1023 are shown in color.
  For context, upper limits are shown, in black, for clusters in the
  Milky Way, M31 and M81 (see the text).}\label{f5}
\end{figure}

From Figure 5, the best radio luminosity constraint for individual
cluster candidates in NGC\,1023 implies an IMBH mass of $M_{\rm BH} <
3.9 \times 10^5~M_\odot$.  The constraints from the stacks of the old
and the intermediate-age clusters in NGC\,1023 correspond to mean IMBH
masses of $\overline{M_{\rm BH}({\rm old})} < 1.4 \times 10^5~M_\odot$
and $\overline{M_{\rm BH}({\rm intermediate~age})} < 2.4 \times
10^5~M_\odot$, respectively.  However, recalling that the clusters'
stellar masses are $M_\star \sim 0.5-40 \times 10^5~M_\odot$, IMBH
constraints are only physically interesting amongst the most massive
clusters with, say, about twice as much mass in stars as in a putative
IMBH.

Applying a z-band cut of 20.5 selects the most massive clusters, all
old and with stellar masses $M_\star \gtrsim 7.5 \times 10^5~M_\odot$
\citep{for14}.  The AIPS task {\tt comb} was used to form Figure 6,
the weighted-mean stack of those massive clusters.  That stack's
3$\sigma$ radio luminosity is $\overline{L_{\rm R}({\rm massive})} <
1.0 \times 10^{34}$ erg s$^{-1}$.  Figure 5 then yields an inferred
mean IMBH mass of $\overline{M_{\rm BH}({\rm massive})} < 2.3\times
10^5~M_\odot$ for the population of massive clusters.  This is a
physically interesting constraint as it suggests black-hole mass
fractions of $\overline{M_{\rm BH}({\rm massive})} / M_\star <
0.05-0.29$ for the massive star clusters.  Cast somewhat differently,
the clusters' mean stellar mass is $1.4 \times 10^6~M_\odot$, implying
a mean black-hole mass fraction of less than 0.16.  Still, it should
be kept in mind that the interpretation of $\overline{M_{\rm BH}({\rm
    massive})}$ is affected by unknowns like the distribution function
of IMBH masses and the fraction of clusters occupied by an IMBH.

\begin{figure}[t]
\plotone{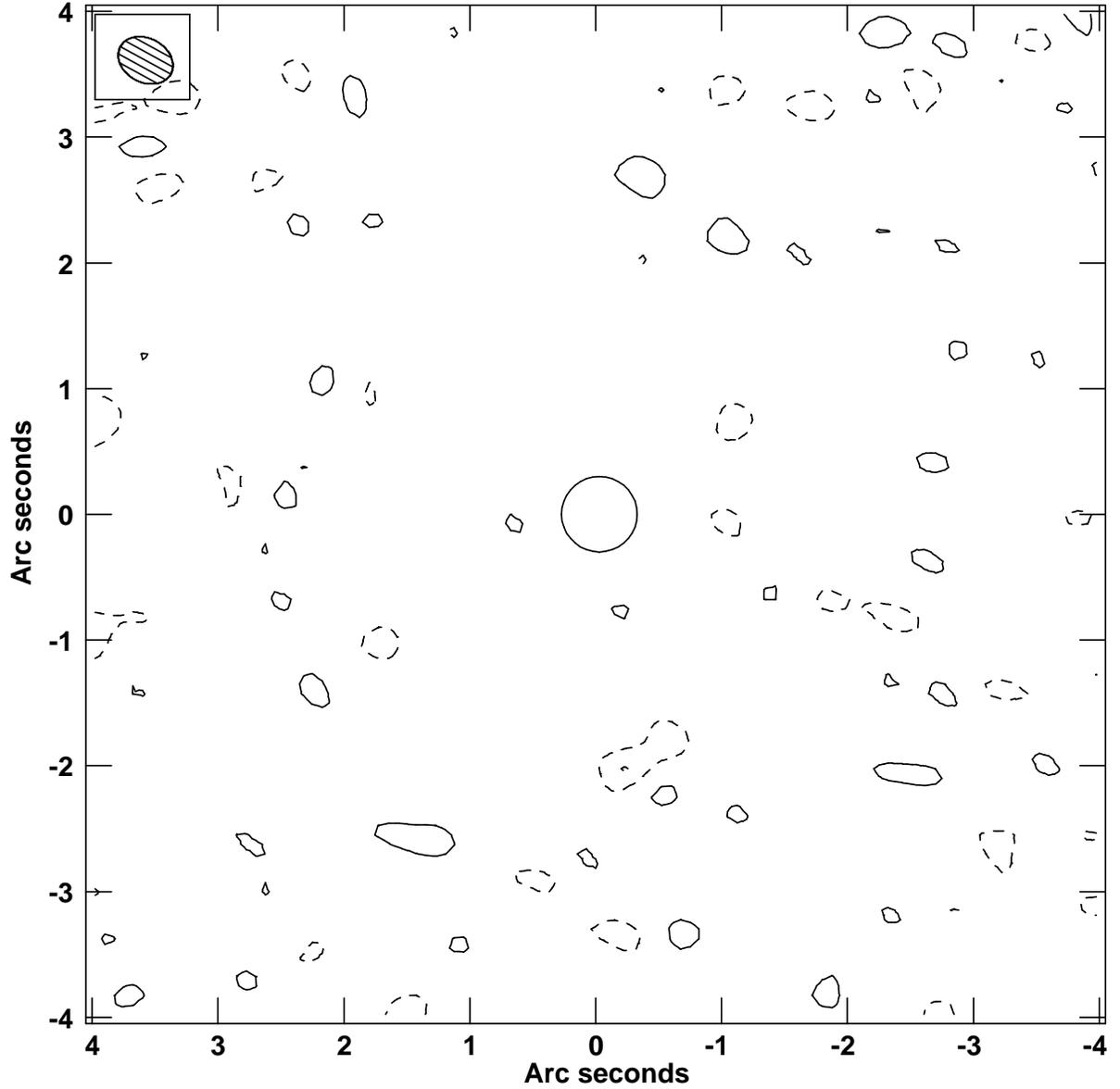}
\caption{Weighted-mean stack of the VLA images of the 20 most massive
  clusters in Figure 1.  The stacked image has an rms noise of
  4.3~$\mu$Jy~beam$^{-1}$ (1$\sigma$).  The hatched ellipse,
  contouring scheme and circle are the same as for Figure 1.  The
  stack measures the mean contribution to the total radio emission
  from the most massive clusters.  No emission is detected above
  3$\sigma$ = 12.9~$\mu$Jy~beam$^{-1}$.}\label{f6}
\end{figure}

Three massive clusters have been spectroscopically confirmed as being
associated with NGC\,1023 \citep{for14} and merit special note.  ID 33
and ID 269 are each as luminous as G1.  Assuming a uniform IMBH mass
distribution and a high IMBH occupation fraction, ID 33 and 269 each
have $\overline{M_{\rm BH}({\rm massive})} / M_\star < 0.05$; for
comparison, the individual values inferred for G1 are $\sim 0.02$ from
a sphere-of-influence approach \citep{geb05}, $< 0.01$ from a
fundamental-plane approach \citep{mil12} and $< 0.02$ from Figure 5.
Under the same assumptions, ID 102, an ultra-compact dwarf, has
$\overline{M_{\rm BH}({\rm massive})} / M_\star < 0.23$; for
comparison, \citet{set14} reported an individual value of 0.15 for the
ultra-compact dwarf M60-UCD1 from a sphere-of-influence approach.
These comparisons suggest that pushing $\overline{M_{\rm BH}({\rm
    massive})}$ lower by factors of a few could reveal analogs of G1
and M60-UCD1 in NGC\,1023.

Deeper VLA imaging of these massive clusters in NGC\,1023 is feasible
but, already, this stacking analysis is sensitive to the mass regime
targeted for dynamical studies of extragalactic star clusters with
next-generation near-infrared telescopes \citep[e.g.,][]{do14}.
Moreover, these and additional massive clusters could be surveyed to a
uniform depth in a future VLA mosaic that covers the full diameter,
$12\farcm6$ (41 kpc), of NGC\,1023's system of about 500 star clusters
\citep{you12,kar14}.  In addition, the methodology demonstrated in
this work on NGC\,1023 could be applied to other nearby early-type
galaxies with rich systems of star clusters \citep[e.g.,][]{bro14}.
Finally, although the conservative, semi-empirical model of
\citet{mac08,mac10} is appealing because of its simplicity, as the
photometric data improve one could consider applying more complex
models with additional free parameters \citep[e.g.,][]{sun13}.

\section{Summary and Conclusions}\label{sumcon}

We used VLA data at 5.5 GHz to search for the radiative signatures of
IMBH accretion from 337 candidate star clusters in NGC\,1023, an
early-type galaxy at a distance of 11.1 Mpc.  None of the individual
clusters were detected.  Similarly, only upper limits were obtained
from weighted-mean image stacks of the 311 clusters with ages older
than 5 Gyr, the 26 clusters with ages of a few hundred Myr, and the 20
clusters with stellar masses $M_\star \gtrsim 7.5 \times
10^5~M_\odot$.  We explored the implications of these data for IMBHs,
if any exist in these clusters.  Our principal findings are as
follows:

\begin{enumerate}
\item The 337 candidate clusters in NGC\,1023 lack radio analogs of
  HLX-1, a strong IMBH candidate in a cluster in the early-type galaxy
  ESO\,243-49.  This suggests that HLX-1 is accreting gas related to
  the formation and/or presence of the 20-Myr-old stars in its host
  cluster.  Alternatively, the HLX-1 phenomenon could be so rare that
  no radio analog is expected in NGC\,1023.
\item Three candidate clusters in NGC\,1023 exhibit hard-state X-ray
  emission.  From the empirical fundamental-plane relation, the
  clusters' X-ray and radio luminosities suggest individual IMBH
  masses of $M_{\rm BH} < 10^6~M_\odot$.  Given the clusters' stellar
  masses this is not a physically interesting constraint.  To make
  progress, a deeper X-ray survey that detects other clusters is
  needed.
\item Using a formalism previously applied to Milky Way star clusters,
  the radio stack of the massive clusters in NGC\,1023 implies an IMBH
  mass of $\overline{M_{\rm BH}({\rm massive})} < 2.3 \times
  10^5~M_\odot$.  This physically interesting constraint can be
  improved with deeper VLA imaging.  Also, more massive clusters could
  be added via a VLA mosaic that fully covers NGC\,1023's star cluster
  system.
\end{enumerate}

\acknowledgments We thank the referee for a helpful report, Dr.~Vince
Pota for estimating the astrometric accuracy of the HST positions
reported in \citet{for14} and Dr.~Craig Walker for advice on image
stacking.  JCAMJ is the recipient of an Australian Research Council
Future Fellowship (FT140101082).  JMW is grateful to the NRAO
Director, Dr.~Tony Beasley, for approving the leave of absence that
facilitated this research.

{\it Facilities:} \facility{VLA}.


\end{document}